\documentclass{aa}
\usepackage[varg]{txfonts}
\usepackage{dblfloatfix}
\usepackage{graphicx}
\usepackage{epstopdf}
\usepackage{pdflscape}
\usepackage{supertabular,booktabs}
\usepackage{txfonts}
\usepackage{gensymb}
\usepackage{color}
\usepackage{url}
\usepackage{multirow}
\usepackage{amsmath}
\usepackage{amstext}
\usepackage{natbib}
\usepackage{longtable}
\usepackage{siunitx}
\usepackage{float}
\usepackage[export]{adjustbox}
\usepackage{kantlipsum}

\begin{document}

\title{On the dust of tailless Oort-cloud comet C/2020 T2 (Palomar) }

\author{Yuna G. Kwon\inst{\ref{inst1}\thanks{Alexander von Humboldt Postdoctoral Fellow}}~\and~Joseph R. Masiero\inst{\ref{inst2}}~\and~Johannes Markkanen\inst{\ref{inst1},\ref{inst3}}}

\institute{Institut f{\" u}r Geophysik und Extraterrestrische Physik, Technische Universit{\" a}t Braunschweig,  Mendelssohnstr. 3, 38106 Braunschweig, Germany\\
\email{y.kwon@tu-braunschweig.de}\label{inst1} 
\and Caltech/IPAC, 1200 E California Blvd, MC 100-22, Pasadena, CA 91125, USA\label{inst2}
\and Max Planck Institute for Solar System Research, Justus-von-Liebig-Weg 3, 37077 G{\"o}ttingen, Germany\label{inst3}}

\date{Received August 31, 2022 / Accepted \today}

\abstract{We report our new analysis of Oort-cloud comet C/2020 T2 (Palomar) (T2) observed at 2.06 au from the Sun (phase angle of 28.$^{\rm \circ}$5) about two weeks before perihelion. T2 lacks a significant dust tail in scattered light, showing a strong central condensation of the coma throughout the apparition, reminiscent of so-called Manx comets. Its spectral slope of polarized light increases and decreases in the $J$ (1.25 $\mu$m) and $H$ (1.65 $\mu$m) bands, respectively, resulting in an overall negative (blue) slope ($-$0.31$\pm$0.14 \% $\mu$m$^{\rm -1}$) in contrast to the red polarimetric color of active comets observed at similar geometries. The average polarization degree of T2 is 2.86$\pm$0.17 \% for the $J$ and 2.75$\pm$0.16 \% for the $H$ bands. Given that near-infrared wavelengths are sensitive to the intermediate-scale structure of cometary dust (i.e., dust aggregates), our light-scattering modeling of ballistic aggregates with different porosities and compositions shows that polarimetric properties of T2 are compatible with low-porosity ($\sim$66 \%), absorbing dust aggregates with negligible ice contents on a scale of 10--100 $\mu$m (density of $\sim$652 kg m$^{\rm -3}$). This is supported by the coma morphology of T2 which has a viable $\beta$ (the relative importance of solar radiation pressure on dust particles) range of $\lesssim$10$^{\rm -4}$. Secular evolution of $r$-band activity of T2 from archival data reveals that the increase in its brightness accelerates around 2.4 au pre-perihelion, with its overall dust production rate $\gtrsim$100 times smaller than those of active Oort-cloud comets. We also found an apparent concentration of T2 and Manx comets toward ecliptic orbits. This paper underlines the heterogeneous nature of Oort-cloud comets which can be investigated in the near future with dedicated studies of their dust characteristics.}


\titlerunning{Dust of Oort-cloud comet C/2020 T2 (Palomar)}

\authorrunning{Y. G. Kwon et al.}

\maketitle

\section{Introduction \label{sec:intro}}

Comets preserve the least-altered planetesimals from the nascent solar system. In particular, comets from the Oort cloud (Oort-cloud comets, OCCs), a reservoir of objects at the outskirts of the solar system ($\sim$10$^{\rm 3-5}$ au from the Sun; \citealt{Dones2015}), spent most of their lives in places where sunlight hardly reaches, making them one of the most important classes of primitive objects to connect their properties back to the early solar system environment. This motivates the upcoming `Comet Interceptor' mission, nominally scheduled for launch in 2029, to explore the early solar system environment via OCCs \citep{Snodgrass2019}.

Observations over the past decades have revealed a heterogeneous nature in OCCs. Departing from the traditional view of comets extending a conspicuous dust tail thousands of kilometers from the nuclei \citep{Krishna Swamy2010}, OCCs in a low-to-moderate activity level (e.g. \citealt{Licandro2019,Garcia2020}) or even without dust tails (so-called Manx comets; \citealt{Meech2016}) have been found, implying a variety of origins for comets found in the present-day Oort
cloud. Given the lack of in-situ data on OCCs, characterization of their dust constituents and context are of particular
importance for a better understanding of their formation and subsequent evolutionary history and support of future space missions.

Here we report a new near-infrared spectropolarimetric observation of Oort-cloud comet C/2020 T2 (Palomar) (hereafter T2) obtained about two weeks before perihelion ($q$ = 2.05 au). Spectropolarimetry provides a degree of linear polarization as a function of wavelength, which is independent of the number density of dust particles but sensitively
depends on their microphysical properties (size, structure, and composition; \citealt{Kiselev2015,Kwon2022}). Together with archival data and modeling of both the light scattered by dust and the dust motion in the coma, we aim to constrain the dust environment of T2 and compare its properties with other OCC observations.

\section{Observations and Data Analysis \label{sec:obsdata}}

\begin{table*}[!h]
\centering
\caption{Geometry and instrument settings of the observations of C/2020 T2 (Palomar)}
\vskip-1ex
\begin{tabular}{c|c|c|c|c|c|c|cccc}
\hline
\hline
Telescope/ & \multirow{2}{*}{Mode} & \multirow{2}{*}{Filter} &  \multirow{2}{*}{Median UT} & \multirow{2}{*}{$N$} & {Exptime} & \multirow{2}{*}{$X$} & $m_{\rm V}$ & $r_{\rm H}$ & $\Delta$ & $\alpha$ \\
Instrument & & & & & (sec) &  & (mag) & (au) & (au) & (\degree) \\
\hline
\hline
\multirow{5}{*}{Hale/} & \multirow{4}{*}{Spol} & \multirow{2}{*}{$J$} & \multirow{2}{*}{04:58:34} & \multirow{2}{*}{32} & \multirow{2}{*}{960} & 1.18 & \multirow{6.5}{*}{13.623} & \multirow{6.5}{*}{2.062} & \multirow{6.5}{*}{1.559} & \multirow{6.5}{*}{28.5} \\
\multirow{4.8}{*}{WIRC$+$Pol} & & & & &  & (1.13--1.22) & & & & \\
\cline{3-7}
 & & \multirow{2}{*}{$H$} & \multirow{2}{*}{04:58:55} & \multirow{2}{*}{32} & \multirow{2}{*}{960} & 1.17 & & & & \\
 & & & & & & (1.15--1.19) & & & & \\ 
\cline{2-7}
 & \multirow{2}{*}{Img} & \multirow{2}{*}{$J$} & \multirow{2}{*}{05:25:08} & \multirow{2}{*}{4} & \multirow{2}{*}{120} & 1.24 & & & & \\
 & & & & & & (1.23--1.24) & & & & \\
 \hline
 \hline
\end{tabular}
\tablefoot{Top headers: Modes, `Spol' and `Img' denote spectropolarimetric and imaging observations, respectively; $N$, number of exposures; Exptime, the total on-source time in seconds (=$N$ $\times$ 30 sec); $X$, average airmass with its range in the bracket; $m_{\rm V}$, apparent total $V$-band magnitude provided by the JPL Horizons (http://ssd.jpl.nasa.gov/?horizons); $r_{\rm H}$ and $\Delta$, heliocentric and geocentric distances in au, respectively; $\alpha$, phase angle (angle of Sun--comet--observer) in degrees.} 
\label{t01}
\vskip-1ex
\end{table*}

A one-epoch low-resolution ($\lambda$/$\Delta \lambda$$\sim$100) near-infrared ($\sim$1.1--1.9 $\mu$m) spectropolarimetric observation of T2 was conducted on UT 2021 June 26.2 using a spectropolarimeter attached to the 5.1-m diameter Hale Telescope at Palomar Observatory (116\degr51$\arcmin$54$\arcsec$W, 33\degr21$\arcmin$23$\arcsec$N, 1 712 m). WIRC$+$Pol is a newly commissioned spectropolarimetry mode of the Wide-field InfraRed Camera (WIRC; \citealt{Wilson2003}) with a field of view of 4\farcm3 $\times$ 4\farcm3 and seeing-limited angular resolution of $\simeq$1\farcs2, located at the prime focus of the telescope \citep{Tinyanont2019}. It measures full linear Stokes parameters ($I$, $Q$, and $U$) as a function of wavelength with one exposure, making resulting datasets free from sky rotation during a sequential exposure of images. A half-wave plate (HWP) rotates the incoming light by four different angles (2$\theta_{\rm HWP}$, where $\theta_{\rm HWP}$ = 0\degree, 22\fdg5, 45\degree, 67\fdg5) that allows for beam-swapping and improved calibration. Each of the four beams then passes a quarter-wave plate, so-called PG (acting as a beam-splitting polarizer and a grating simultaneously), and an opaque mask in a row, thereby leaving their trace in four quadrants on the CCD, 3\arcmin\ away from one another (Fig. 1 in \citealt{Tinyanont2019}). We obtained 32 dithered images (16 each in slit A and B positions) in the $J$ and $H$ bands to subtract background signals. More details of the observing strategy can be found in \citet{Masiero2022}. The observation journal is summarized in Table \ref{t01}.

Basic calibration, spectral extraction, and polarimetric calculation were all performed in the WIRC$+$Pol Data Reduction
Pipeline\footnote{\url{https://wircpol.readthedocs.io/en/latest/}}, whose details are described in \citet{Tinyanont2019}. An additional correction was made to residual offsets in polarization angle by adding 6\fdg101 and 3\fdg416 to the $J$- and $H$-band values, respectively, based on Figure 3 of \citet{Masiero2022}. In the pipeline, Stokes parameters were corrected from the instrumental polarization and polarization efficiency.  We extracted spectra of each resolution element in three aperture sizes, corresponding to 850 km, 1 140 km, and 1 420 km in cometocentric distances at the time of our observation, yet the latter of which was discarded due to the significant background emission at $\gtrsim$1.7 $\mu$m. To increase the signal-to-noise ratio, we smoothed the extracted spectra over five spectral bins and eliminated the measurements deviating by more than 5$\sigma$. The resulting polarimetric parameters (the degree of linear polarization and its position angle) were transformed into the scattering plane (a plane containing the Sun--comet--Earth) in the same manner as \citet{Chernova1993}.

\section{Results and Discussion \label{sec:res}}

\begin{figure}[!t]
\vskip-1ex
\centering
\includegraphics[width=8cm]{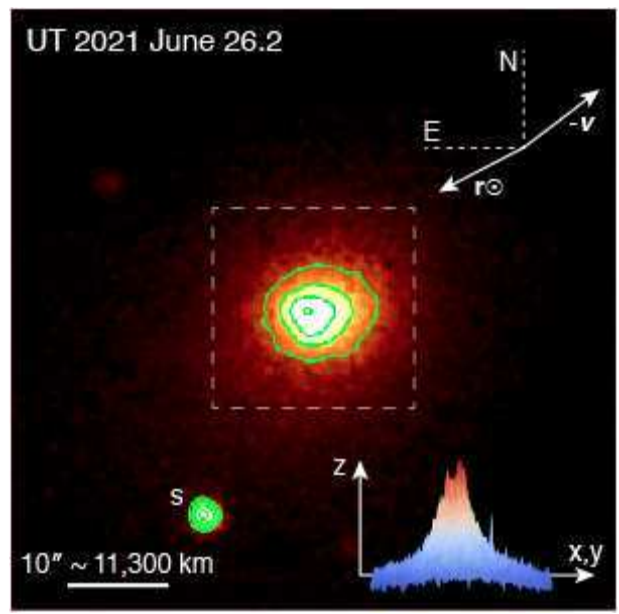}
\caption{$J$-band composite image of T2. The image was boxcar smoothed in the 3-pixel width and overlapped with contours at 90, 50, 25, and 5 \% of T2's peak brightness. `S' marks a background star. A radial profile is given over the inner coma region (enclosed by a dashed-line square of 20\arcsec\ by 20\arcsec) in the lower right corner. The negative velocity  ($-$${\bf{\vec{v}}}$) and solar radius (${\bf{\vec{r_{\rm \odot}}}}$) vectors are given.}
\label{Fig01}
\vskip-1ex
\end{figure}

Figure \ref{Fig01} shows a $J$-band composite image of T2. A featureless, spherical coma is notable, distinct from the morphology of typical OCCs (e.g. \citealt{Bauer2015}). Most of the coma signal comes from the central part within $\sim$5\arcsec\ from the photocenter. Similar trends throughout the apparition of the comet are confirmed from $r$-band archival data (Appendix \ref{sec:app1}).

\subsection{Polarimetric properties of the coma dust}

\begin{figure}[!t]
\vskip-1ex
\centering
\includegraphics[width=9cm]{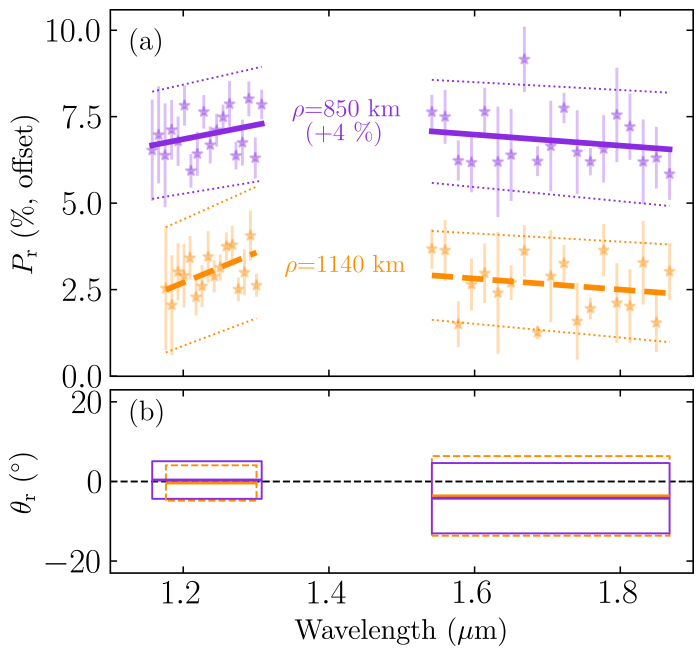}
\caption{Degree of linear polarization of the dust of T2 $P_{\rm r}$ and its position angle $\theta_{\rm r}$ at a phase angle of 28\fdg5 are given as a function of wavelength $\lambda$ in Panels a and b, respectively. $P_{\rm r}(\lambda)$ extracted from different aperture sizes ($\rho$) are offset for clarity. In Panel a, datapoints in each $J$- and $H$-band region were fitted by a linear least-square function (given as thick solid and dashed lines), where upper and lower dotted lines indicate 1$\sigma$ intervals. In Panel b, the 1$\sigma$ regions of $\theta_{\rm r}$ and their central values are given.}
\label{Fig02}
\vskip-1ex
\end{figure}
\begin{figure}[!b]
\vskip-1ex
\centering
\includegraphics[width=9cm]{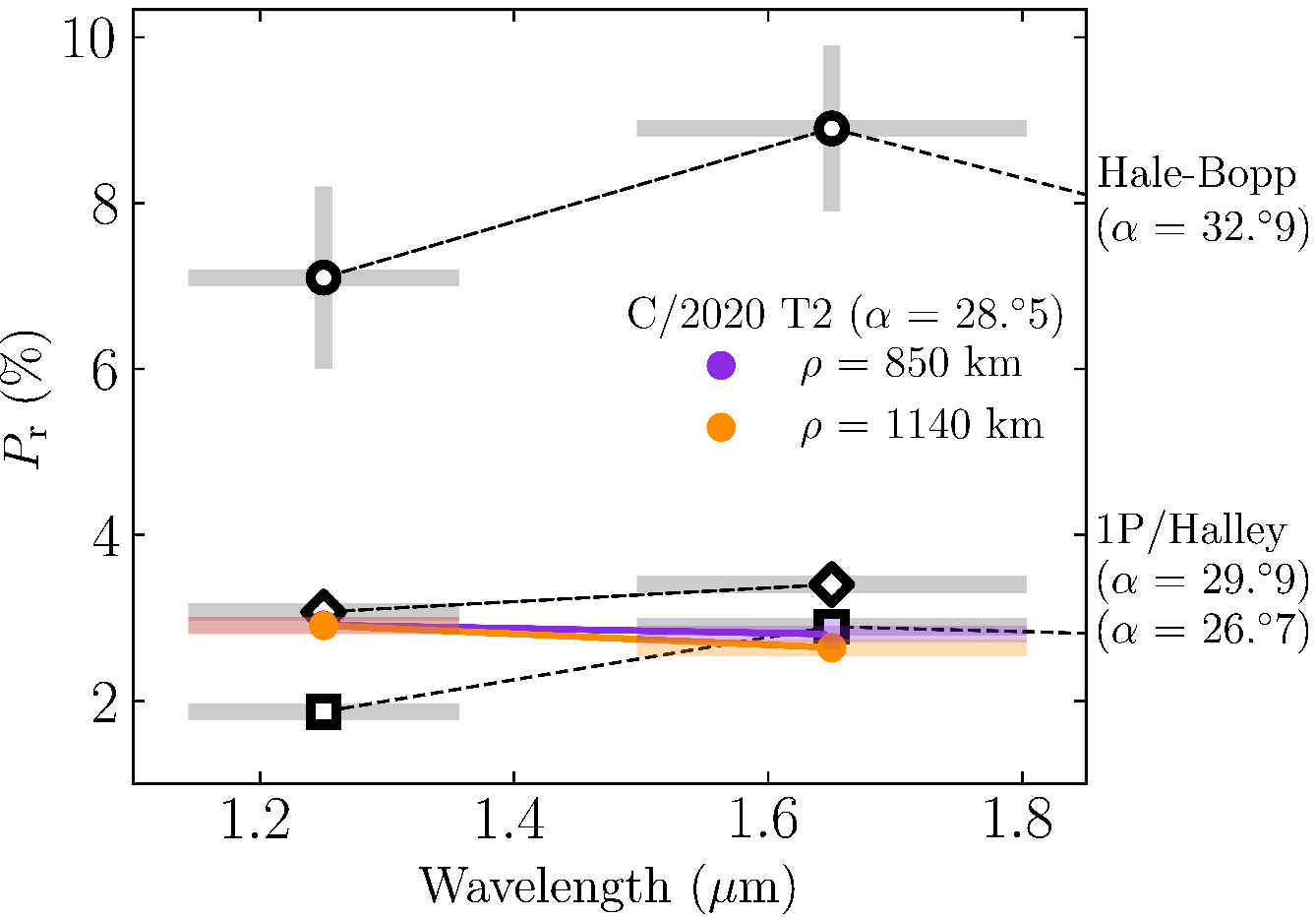}
\caption{Spectral dependence of the $P_{\rm r}$ of T2 and other OCCs: C/1995 O1 (Hale-Bopp) and 1P/Halley. Measurements for other OCCs were obtained from the NASA/PDS comet polarimetric archive \citep{Kiselev2017}: Hale-Bopp from \citet{Hasegawa1997} and 1P/Halley from \citet{Brooke1987} (squares) and \citet{Kikuchi1987} (diamonds), whose phase angles are given in parentheses. The horizontal bars cover the $J$ and $H$ bands, while the vertical bars indicate errors in the average value of all data points in each band. }
\label{Fig03}
\vskip-1ex
\end{figure}

Figure \ref{Fig02} shows the degree of linear polarization $P_{\rm r}$ and its position angle with respect to the normal direction of the scattering plane $\theta_{\rm r}$ extracted from two different aperture sizes $\rho$ as a function of wavelength $\lambda$. The 1$\sigma$ intervals of $\theta_r$ are distributed close to zero, which precludes significant dust alignment over the region analyzed and supports the reliability of our data reduction. The two spatial resolutions have almost the same $P_{\rm r}$ values at a given wavelength and similar spectral dependence: a slight increase in the $J$ band (red polarimetric color; 4.13$\pm$2.11\% $\mu$m$^{\rm -1}$ for $\rho$ = 850 km and 8.68$\pm$3.01 \% $\mu$m$^{\rm -1}$ for $\rho$ = 1 140 km) and inflection point around 1.4 $\mu$m, followed by a decrease in the $H$ band (blue polarimetric color; $-$1.59$\pm$1.99 \% $\mu$m$^{\rm -1}$ for $\rho$ = 850 km and $-$1.58$\pm$1.72 \% $\mu$m$^{\rm -1}$ for $\rho$ = 1 140 km). There is a local peak of $P_{\rm r}$ ($\approx$1.5$\sigma$ significance) at $\sim$1.65 $\mu$m for the inner coma ($\rho$ = 850 km) data. This region corresponds to an absorption peak of crystalline water-ice \citep{Grundy1998} hosted by a broader 1.5-$\mu$m band that has been observed for several active comets (e.g. \citealt{Kawakita2004}). We tested different smoothing parameters, but the feature is always present and becomes statistically indistinguishable in the larger aperture size. Although polarimetry is sensitive to the rapid change of the refractive index of scattering materials, in the absence of data covering the region where deeper bands of water ice are expected to be (1.5 and 2.0 $\mu$m; \citealt{Mastrapa2008}), we will not consider the weak 1.65 $\mu$m peak in the following analysis.

The polarimetric color $P_{\rm r}(\lambda)$ of T2 is then compared with those of other OCCs observed at similar geometries (Fig. \ref{Fig03}). We selected archival data for which observations at multiple wavelengths were conducted simultaneously (or at least on the same night). For comparison, representative $P_{\rm r}$ values of T2 were used by averaging data points in each band: 2.91$\pm$0.15 \% for the $J$ and 2.84$\pm$0.14 \% for the $H$ bands over $\rho$ = 850 km; and 2.80$\pm$0.19 \% for the $J$ and 2.62$\pm$0.18 \% for the $H$ bands over $\rho$ = 1 140 km. Since the two apertures offer consistent values within the errors, we averaged the $P_{\rm r}$ at the same band and derived the overall polarimetric color as $-$0.31$\pm$0.14 \% $\mu$m$^{\rm -1}$.  T2's average $P_{\rm r}$ values themselves are comparable to those of 1P/Halley, which seems natural for cometary dust in this small $\alpha$ region, as $P_{\rm r}$ begins to show discernible deviations between comets at larger $\alpha$ ($\gtrsim$40\degree; \citealt{Kiselev2015}), except for the exceptionally high $P_{\rm r}$ of Hale-Bopp \citep{Hadamcik1997}. However, the blue $P_{\rm r}(\lambda)$ of T2 is in contrast to the dust of OCCs that generally exhibit red polarimetric color over the $J$ and $H$ bands.

T2's lower $H$-band $P_{\rm r}$ as a result of the blue $P_{\rm r}(\lambda)$ indicates that its dust has a heterogeneity larger than other cometary dust at the given wavelength \citep{Bohren1983}. This can be achieved by its compositional (lower absorptivity, e.g. lower fractions of amorphous carbon or abundant silicates; \citealt{Rouleau1991,Greenberg1996}) and/or mechanical aspects (e.g. higher packing density; \citealt{Kolokolova2010}) to consequently enhance electromagnetic interactions among constituting grains. Compact water ice larger than a few tens of micrometers yields blue $P_{\rm r}(\lambda)$ but also increases overall $P_{\rm r}$ values \citep{Warren1984,Warren2019} and so is not consistent with T2. If the dust is small (of order $\sim$0.1--1 $\mu$m) or fluffy (porosity as low as 2 \%) it will behave similarly to individual constituting grains in dynamics \citep{Mukai1992,Skorov2016} and in light scattering \citep{Kolokolova2011}.  It will then provide a red $P_{\rm r}(\lambda)$ similar to that of 1P/Halley and Hale-Bopp (Fig. \ref{Fig03}).  In this case its enhanced thermal emission close to the Sun ($r_{\rm H}$ $\lesssim$1 au) can depolarize signals longward of the $H$ band \citep{Oishi1987}, but this would not be significant for our T2 observations at 2.06 au from the Sun. Together with the coma morphology (Fig. \ref{Fig01}), it is evident for T2 that the observed polarimetric behaviors of its coma cannot be explained by the typical dust properties used to characterize observations of active OCCs.

\subsection{Quantitative estimations of the optical and dynamical properties of the T2 dust}

\begin{figure*}[!t]
\vskip-1ex
\centering
\includegraphics[width=0.7\textwidth]{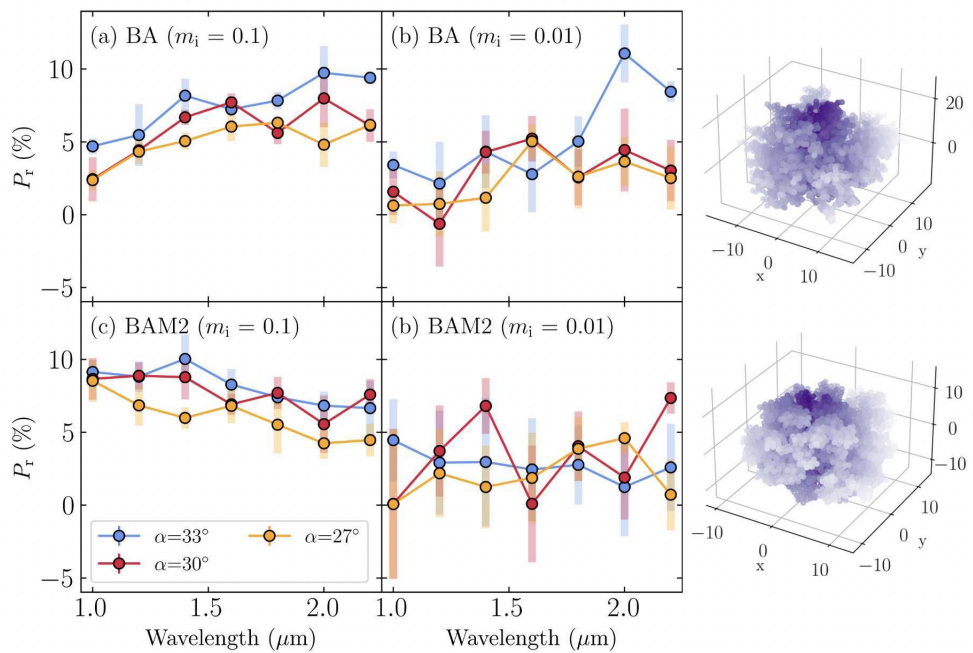}
\caption{Orientation- and realization-averaged $P_{\rm r}$ as a function of wavelength. The upper and lower rows show results for BA and BAM2, respectively, whose geometry is visualized on the rightmost side of the micrometer scale. The left columns display outputs in the standard composition ($m_{\rm r}$ = 1.6 and $m_{\rm i}$ = 0.1), while the right columns show those with a lower absorptivity ($m_{\rm i}$ = 0.01). Errors indicate 1$\sigma$ for the four realizations of the clusters. 
}
\label{Fig04}
\vskip-1ex
\end{figure*}

To reproduce the blue polarimetric color $P_{\rm r}(\lambda)$ of T2, we conducted light-scattering modeling of dust aggregates. Presuming that cometary dust is hierarchical \citep{Guttler2019}, a $>$100-$\mu$m dust agglomerate consists of $\sim$10--100-$\mu$m aggregates, each of which is in turn composed of $\sim$0.1 $\mu$m grains (monomers). In this regard, near-infrared wavelengths are most sensitive to the intermediate-scale structure (i.e., dust aggregate). We thus considered two types of ballistic aggregates in the same fractal dimension ($\approx$3)\footnote{Rosetta observations of coma dust showed particle strength can vary across a large agglomerate, where smaller dust constituents ($\sim$15--40 $\mu$m)  have higher strength \citep{Hornung2016}. This makes us exclude fluffy dust (fractal dimension of 1.5--2.5; \citealt{Guttler2019}), which is relevant to a $\sim$millimeter-sized parent dust \citep{Fulle2015,Mannel2016}, from consideration.} and mass but in different porosities \citep{Shen2008}: ballistic aggregate (BA, porosity $\mathcal{P}$$\approx$87 \% and characteristic cluster radius $R$$\sim$15.8 $\mu$m) and BA with two migrations (BAM2, $\mathcal{P}$$\approx$66 \% and $R$$\sim$11.5 $\mu$m). The {\it Rosetta} mission to comet 67P/Churyumov-Gerasimenko showed that monomer size is distributed over $\sim$0.05--0.6 $\mu$m but weighted toward smaller size with a mean of $\sim$0.1 $\mu$m \citep{Mannel2019}. Due to computational limitations, however, we cannot probe a large aggregate of $\lesssim$0.1-$\mu$m monomers. \citet{Shen2008,Shen2009} verified that the size of monomers would be secondary as long as they are smaller than the wavelength considered; thus, we considered a simple case of dust clusters consisting of 4 096 spherical monomers of 0.5-$\mu$m in radius\footnote{We expect $P_{\rm r}$ to be independent of the size for larger aggregate scales of interest in typical Halley-dust composition \citep{Mackowski2022}. This makes us consider a simple dust cluster of monodisperse monomers.}. Their geometry \citep{Shen2008}\footnote{\url{https:// www.astro.princeton.edu/~draine/agglom.html}} was randomly generated and averaged over four realizations. 
For each dust realization, we defined 64 scattering planes (each plane has a constant azimuthal angle $\phi$) and then averaged the outputs. We repeated these processes at 128 random orientations of each dust cluster and averaged the results.
We also compared two composition cases: an average complex refractive index of $m_{\rm r}$ = 1.6 and $m_{\rm i}$ = 0.1, which is in the range of the typical cometary dust (e.g. \citealt{Moreno2018}), and slightly more transparent one ($m_{\rm i}$ = 0.01). The 4 $\times$ 4 Mueller matrix was calculated using the fast superposition T-matrix method (FaSTMM;
\citealt{Markkanen2017}).

Figure \ref{Fig04} shows orientation- and realization-averaged $P_{\rm r}$ of the modeled dust as a function of wavelength. In the case of the average composition ($m_{\rm i}$ = 0.1), BA has red $P_{\rm r}(\lambda)$, whereas less-porous BAM2 steadily yields blue $P_{\rm r}(\lambda)$. This $\mathcal{P}$--$P_{\rm r}(\lambda)$ relationship is in line with previous studies of dust-rich comets whose high $P_{\rm r}$ and red $P_{\rm r}(\lambda)$ are associated with the
presence of highly porous particles in the coma and for Hale-Bopp with extremely fluffy aggregates (\citealt{Kiselev2015} and references therein). In the case of less absorbing dust ($m_{\rm i}$ = 0.01), BA shows slight red $P_{\rm r}(\lambda)$ despite large fluctuations due to enhanced interparticle scattering which contributes to an unstable trend\footnote{In part, resonance effects inside monomers due to the monomer size considered here could contribute to the enhancement of fluctuation.}, while BAM2 gives either inconsistent or a comparably shallow slope of $P_{\rm r}(\lambda)$ to our results. In the absence of other information on T2, this leads us to prefer the standard composition. Slightly higher $P_{\rm r}$ values of BAM2 than the observation might not be critical because in reality scattering dust should be ensembles of aggregates distributed in size, structure, and composition which contain larger heterogeneities than our models and could decrease resulting $P_{\rm r}$ \citep{Bohren1983}. We do not claim that our BAM2 dust is unique, but it can reproduce the characteristic blue $P_{\rm r}(\lambda)$ of T2.  Our results thus suggest that the optical properties of the coma dust of T2 are compatible with BAM2-like lower-$\mathcal{P}$ ($\sim$66 \%) dust aggregates, deficient in higher $\mathcal{P}$ ($\gtrsim$87 \%) ones that are used to describe active comets.

Next, we focus on the coma morphology of T2. At the region of $\gtrsim$1 000 km from the nucleus, dust trajectories in the coma result from a tug-of-war between outward solar radiation pressure and inward solar gravity forces \citep{Finson1968a,Burns1979}. Their mutual effect is parameterized by $\beta$, which is the ratio of the former to the latter:
\begin{equation}
\beta \equiv \frac{F_{\rm rad}}{F_{\rm grav}} = \frac{3 L_{\rm \odot} Q_{\rm pr}}{16 \pi G M_{\odot} c (a \rho_{\rm d})} \sim K\cdot(a \rho_{\rm d})^{\rm -1}
~,
\label{eq:eq1}
\end{equation}
\noindent where $L_{\rm \odot}$ is the solar luminosity, $Q_{\rm pr}$ is the dimensionless coefficient of radiation pressure ($\approx$1), $G$ is the gravitational constant, $M_{\odot}$ is the solar mass, $c$ is the speed of light, and $a$ and $\rho_{\rm d}$ are the radius and density of the dust. Assuming spherical dust with uniform $\rho_{\rm d}$, $K$ is $\sim$ 5.71 $\times$ 10$^{\rm -4}$ kg m$^{\rm -2}$ and $\beta$ becomes a direct function of $a$ and
$\rho_{\rm d}$ (which is pertinent to $\mathcal{P}$). With Equation \ref{eq:eq1}, we can specify in each coma region the $\beta$ of dust having different physical characteristics (syndyne) and ejection times (synchrone). From $r$-band archival images showing the onset of discernible activity for T2 began around 3.1 au (Appendix \ref{sec:app1}), we set a simple model with zero ejection velocity where the dust ejection started $\sim$180 days prior to our observation. Additional forces that can alter dust motions in the coma (e.g. sublimation and fragmentation) are not
considered here.

\begin{figure*}[!t]
\vskip-1ex
\centering
\includegraphics[width=\textwidth]{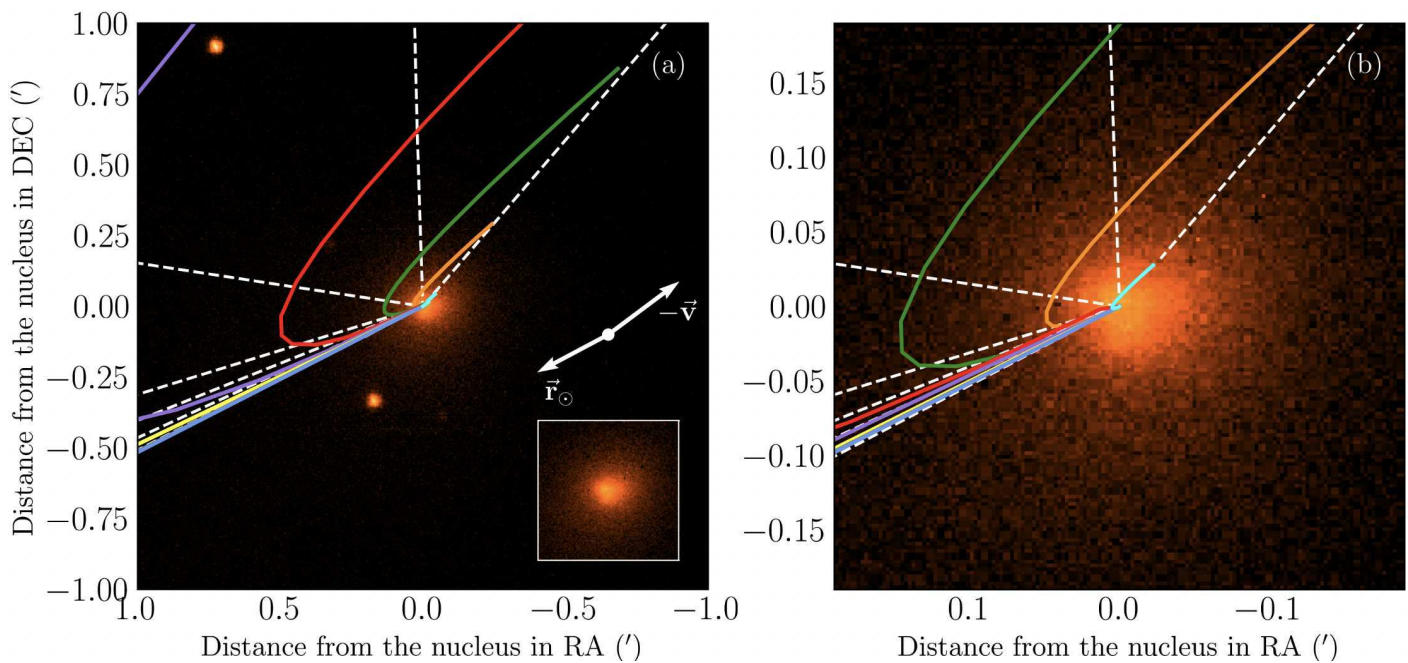}
\caption{(a) Synchrones and syndynes of the T2 coma on UT 2021 June 26.2. The dashed lines are synchrones, indicating the locations of dust ejected at different times prior to the observation: 180, 120, 90, 60, 45, 30, 15, and 5 days from right to left. The solid curves are syndynes, where each has a constant $\beta$ varying from 10$^{\rm -4}$ (low-mobility dust) to 1 (high-mobility dust) anticlockwise from the rightmost. A close-up image is given on the bottom right. (b) Same as T2 in panel a but five times magnified image.}
\label{Fig05}
\vskip-1ex
\end{figure*}

Figure \ref{Fig05} compares modeled synchrones and syndynes with the T2 image that is the same as Figure \ref{Fig01} but with a twice larger field of view. Only the $\beta$ = 10$^{\rm -4}$ syndyne reproduces the compact central part of the coma. Its featureless morphology suggests that dust particles in the coma were ejected over a wide range of times. The small working $\beta$ of $\sim$10$^{\rm -4}$ is in accordance with dust that makes up dust trails ($\sim$mm-sized dust
with $\beta$ $\lesssim$10$^{\rm -4}$; \citealt{Ishiguro2007}) and thus implies that T2 accommodates dust that is less sensitive to solar radiation pressure in its near-nucleus region. This is certainly not typical for the majority of OCCs whose significant dust tails tend to have larger $\beta$ \citep{Moreno2022}.

In tandem with our light-scattering modeling showing that viable dust in the T2 coma has $\mathcal{P}$ of $\sim$66 \% on the scale of a dust aggregate (order of $\sim$10--100 $\mu$m; \citealt{Guttler2019}) in a typical range of the dust composition of comets \citep{Levasseur-Regourd2018}, if we assume that the ice-free dust consists of amorphous carbon (75 vol.\%) and Mg-rich silicates (25 vol.\%), its approximate density ($\rho_{\rm d}$ = $f_{\rm C}$$\rho_{\rm C}$ + $f_{\rm Si}$$\rho_{\rm Si}$, where $f_{\rm C}$ and $f_{\rm Si}$ are the volume fractions of amorphous carbon and Mg-rich silicates, respectively, normalized as $f_{\rm C}$ + $f_{\rm Si}$ + $\mathcal{P}$ = 1) is $\sim$652 kg m$^{\rm -3}$. Here we use densities $\rho_{\rm C}$ = 1 435 kg m$^{\rm -3}$ \citep{Jager1998} and $\rho_{\rm Si}$ = 3 360 kg m$^{\rm -3}$ \citep{Dorschner1995} and $\mathcal{P}$ of 66 \%. Substituting the retrieved density in to $\rho_{\rm d}$ in Eq. \ref{eq:eq1}, $\beta$ $\sim$ 10$^{\rm -4}$ corresponds to the dust size of $a$ $\sim$ 270 $\mu$m. However, we
should be cautious in extrapolating our $\mathcal{P}$ of dust aggregates to larger dimensions since cometary dust is likely hierarchical structures of heterogeneous dust aggregates rather than a homogeneous conglomerate of sub-micrometer grains \citep{Skorov2018,Blum2022}.

\subsection{Possible relationships between the dynamical \& dust properties of Oort-cloud comets}

Finally, to search for a relationship between the dust properties and orbital distribution of OCCs, information diagnosing their dust characteristics was gleaned from previous observations made in polarimetry and/or mid-infrared spectroscopy. Following the criteria suggested by \citet{Kolokolova2007} and \citet{Kwon2021}, we classify comets as `Type II' when they have 1) higher $P_{\rm r}$ than the average trend of OCCs at given $\alpha$, 2) red $P_{\rm r}(\lambda)$, and/or 3) strong intensity of a 10-$\mu$m silicate emission feature ($\gtrsim$1.5; \citealt{Hanner2004}), indicative of high-porosity coma dust; otherwise, they are classified as `Type I' that predominantly eject T2-like lower-porosity dust. Comets with a low-intensity ratio of the C$_{\rm 2}$ emission feature to the dust continuum ($<$500; \citealt{Krishna Swamy2010}) and a sharp anti-solar dust tail are also grouped as Type II. Six Manx-comet candidates are classified in Type I as their (nearly) tailless morphology akin to T2 indicates that they presumably lack high-porosity dust in the coma \citep{Meech2016}. We only considered comets whose perihelion distance $q$ is less than 3.1 au for this classification{\footnote{As OCCs with $q$ $>$ 3.1 au are located outside the water ice line \citep{Blum2014,Gundlach2020}, their relatively low activity and lack of dust features shaped by sufficient solar radiation pressure make the classification criteria adopted here hardly applicable. In addition, given that these distant comets are likely to undergo a rapid change of their $q$ by planet perturbations \citep{Fouchard2017,Vokrouhlicky2019}, we anticipated that their current orbital elements would be less informative.}. As a result, a total of 43 OCCs (= 20 Type I $+$ 23 Type II) are selected. Their orbital elements, types, and references are tabulated in Table \ref{ta1}.

\begin{figure*}[!t]
\vskip-1ex
\centering
\includegraphics[width=\textwidth]{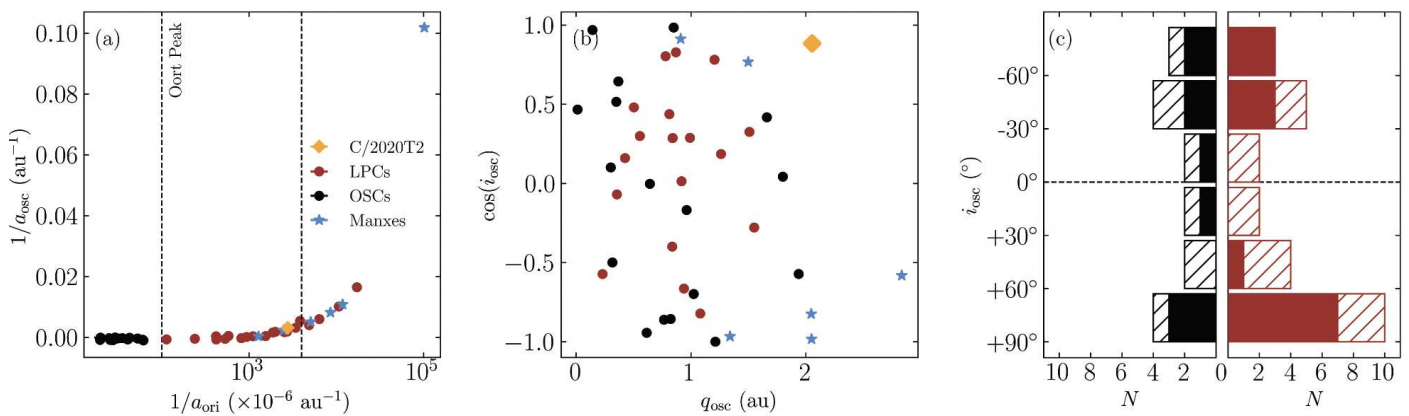}
\caption{(a) Distribution of the reciprocal original semimajor axis 1/$a_{\rm ori}$ and reciprocal osculating semimajor axis 1/$a_{\rm ori}$ of 43 selected OCCs. The Oort Spike ($\sim$10$^{\rm 4}$ au) and the boundary ($\sim$250 au) of planetary regions are marked as the left and right vertical lines, respectively. (b) Distribution of the comets in the plane of the osculating perihelion distance $q_{\rm osc}$ and inclination $\cos(i_{\rm osc})$. (c) Distribution of the comets grouped in 30\degree\ inclination bins. OSCs and LPCs are shown in the left and right panels, respectively, therein the hatched and filled blocks denote Type I and Type II comets. Detailed classification criteria are explained in the text.}
\label{Fig06}
\vskip-1ex
\end{figure*}

Figure \ref{Fig06}a shows a distribution of the reciprocal original semimajor axes 1/$a_{\rm ori}$ that has a definite correlation with the number of approaches to the Sun \citep{Everhart1972} and the reciprocal osculating semimajor axes 1/$a_{\rm osc}$ of the 43 OCCs. The 1/$a_{\rm ori}$ values are quoted from the Warsaw Catalogue of near-parabolic comets \citep{Krolikowska2014}, Minor Planet Center database search engine\footnote{\url{https://www.minorplanetcenter.net/db_search/}} or Nakano note\footnote{\url{https://www.oaa.gr.jp/~oaacs/nk.htm}}. The comets are divided into two dynamical groups: Oort-spike comets (OSCs, black circles) inside the classical Oort Peak at $a_{\rm ori}$ $\sim$ 10$^{\rm 4}$ au \citep{Oort1950} and Long-period comets (LPCs, brown circles) outside the peak. The former is often assumed to be dynamically new, though recent studies suggest that the fiducial line of classification should be located farther than the classical one \citep{Dybczynski2015}. Another boundary is marked at $a_{\rm ori}$ $\sim$ 250 au, inside of which the effect of resonances from Neptune outweigh the Oort-cloud processes (e.g. perturbations of passing stars and the Galactic tide; \citealt{Brasser2012,Fouchard2017}). The $a_{\rm ori}$ of the Manxes and T2 are inside the inner edge of the Oort-cloud at $\sim$1 500 au \citep{Vokrouhlicky2019}.

OSCs and LPCs distribute rather homogeneously in the plane of the current perihelion $q_{\rm osc}$ and inclination $\cos(i_{\rm osc})$ (Fig. \ref{Fig06}b). For Type I (filled black) and Type II (hatched black) comets, there are about 1.5 times as many Type I comets in prograde orbits (Fig. \ref{Fig06}c), though the number of comets per bin here would hardly be related to the real architecture of the Oort cloud since we selected comets only for which decent dust analysis has been made. Nonetheless, the relatively high fraction of Type I comets among the considered ones within $\pm$30\degree\ from the ecliptic plane, particularly the clustering of all Manx comets and T2 toward the plane, is noteworthy. This apparent clustering in space with their similar 1/$a_{\rm ori}$ points might reflect their unique origins
\citep{Meech2016} or evolutionary pathways, different from the majority of OCCs ejected from the region of ice-giant planets \citep{Vokrouhlicky2019}. More observations are required to exclude possible observational bias (e.g. less active comets might be easier to be detected in low relative velocity space to the Earth?). It will be an interesting topic for future studies to see whether this trend can be retained in larger datasets and thus whether it is more probable to observe less-evolved OCCs near perpendicular to the ecliptic plane.

\section{Conclusions \label{sec:concl}}

This paper reports our tailless Oort-cloud comet C/2020 T2 (Palomar) analysis. The main results are as follows.

\begin{enumerate}

\item The $J$-band (1.25 $\mu$m) image of T2 on UT 2021 June 26.2 exhibits tailless morphology, where more than 95 \% of light is concentrated in $\lesssim$10$^{\rm 4}$ km from the nucleus center, reminiscent to the coma morphology of Manx comets \citep{Meech2016}. Secular evolution of its $r$-band activity (Appendix \ref{sec:app1}) supports the overall low activity of the comet.

\item Average $P_{\rm r}$ of T2 at $\alpha$ = 28\fdg5 is 2.91$\pm$0.15 \% for the $J$ and 2.84$\pm$0.14 \% for the $H$ bands over $\rho$ = 850 km; and 2.80$\pm$0.19 \% for the $J$ and 2.62$\pm$0.18 \% for the $H$ bands over $\rho$ = 1 140 km. The aperture-averaged polarimetric color over the $J$--$H$ bands is blue ($-$0.31$\pm$0.14 \% $\mu$m$^{\rm -1}$), opposite to the red polarimetric color of active comets observed at similar $\alpha$.

\item From our light-scattering modeling of ballistic aggregates distributed in compositions and porosities and dust dynamic modeling, we suggest that the coma dust of T2 in the near-infrared is compatible with low-porosity ($\sim$66 \%) dust with typical dust composition ($\rho_{\rm d}$ $\sim$ 625 kg m$^{\rm -3}$). If we assume the uniform distribution in density over the 10--100 $\mu$m aggregate scale, the constrained $\beta$ $\sim$ 10$^{\rm -4}$ corresponds to the viable dust aggregate size of $\sim$270 $\mu$m.

\item We found that Manx comets and T2 share dynamical properties to some extent, showing clustering near the ecliptic plane. It appears that a higher percentage of Type I comets occurs within 30\degree\ of the plane, which needs to be confirmed with more observations.

\end{enumerate}
\begin{acknowledgements}

Y. G. K. gratefully acknowledges the support of the Alexander von Humboldt Foundation. J. M. acknowledges funding from the European Union’s Horizon 2020 research and innovation program under grant agreement No 75390 CAstRA. This paper is based in part on observations obtained at the Hale Telescope, Palomar Observatory as part of a continuing collaboration between the California Institute of Technology, NASA/JPL, Yale University, and the National Astronomical
Observatories of China.

\end{acknowledgements}


\clearpage

\begin{appendix}

\section{$r$-band photometry of C/2020 T2 (Palomar)\label{sec:app1}}

\begin{figure*}[!t]
\centering
\includegraphics[width=0.9\textwidth]{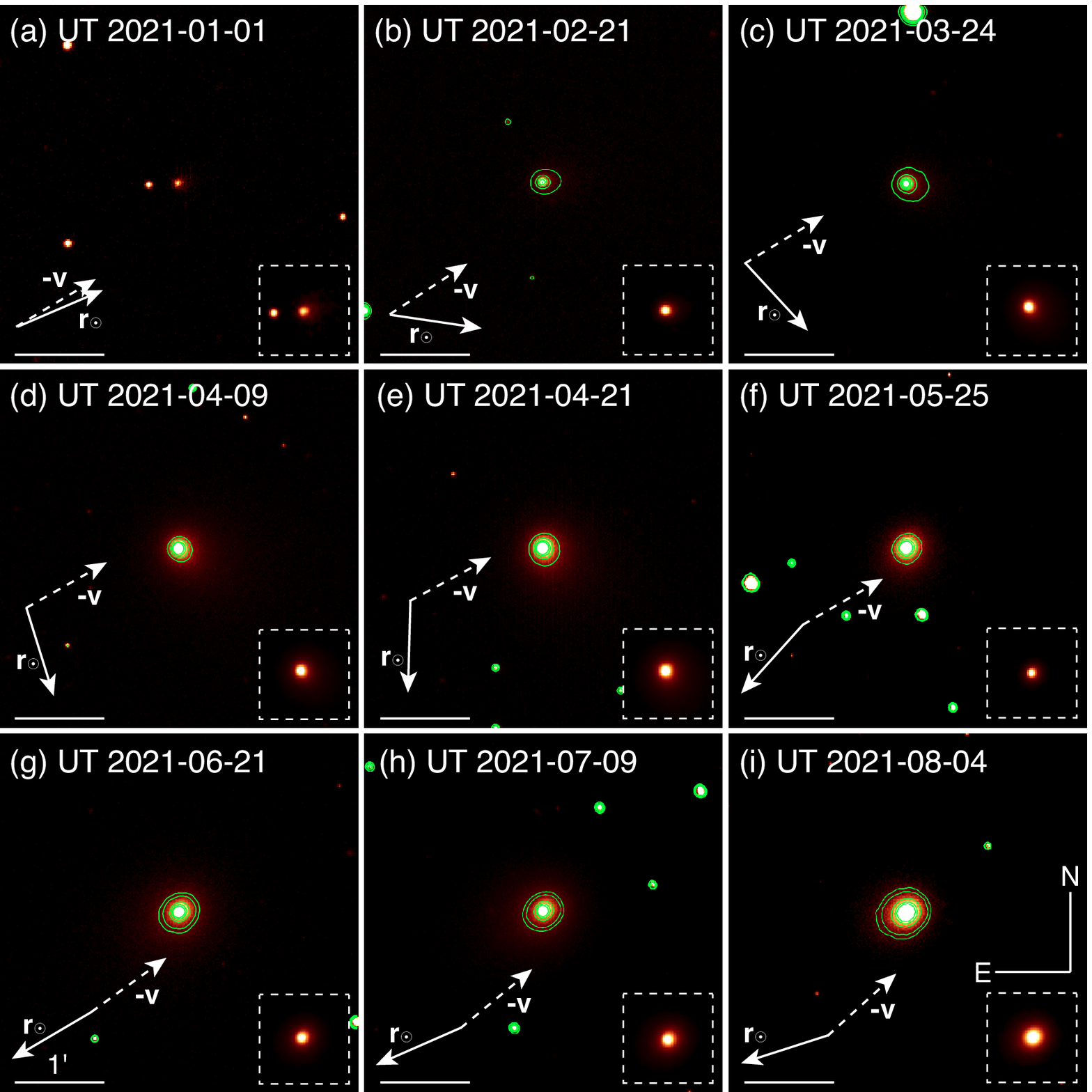}
\caption{Coma morphology of T2 in its inbound (Panels a to h) and outbound (Panel i) orbit taken from the ZTF archive. All panels cover a FoV of 4\arcmin\ $\times$ 4\arcmin\ and provide a close-up image on the bottom right with FoV of 1\arcmin. The negative velocity vector ($-${\bf{\vec{v}}}), solar radius vector ({\bf{\vec{r_\odot}}}), and 1\arcmin\ scale bar are given in each panel. A zoom-in image was processed by the boxcar smoothing with a width of 3 pixels to better visualize coma features. Five contour levels stratify the brightness within 2-$\sigma$ of the peak brightness on a logarithmic scale, except for the image on Panel a where we could not make aperture photometry due to its weak contrast to the background signal. The heliocentric distance for the comet in each panel is (a) 3.004 au (inbound), (b) 2.624 au (inbound), (c) 2.422 au (inbound), (d) 2.329 au (inbound), (e) 2.267 au (inbound), (f) 2.129 au (inbound), (g) 2.068 au (inbound), (h) 2.055 au (inbound, two days before perihelion), and (i) 2.075 au (outbound). North is up and east is to the left.
}
\label{Figap01}
\end{figure*}

\begin{figure}[!t]
\centering
\includegraphics[width=9cm]{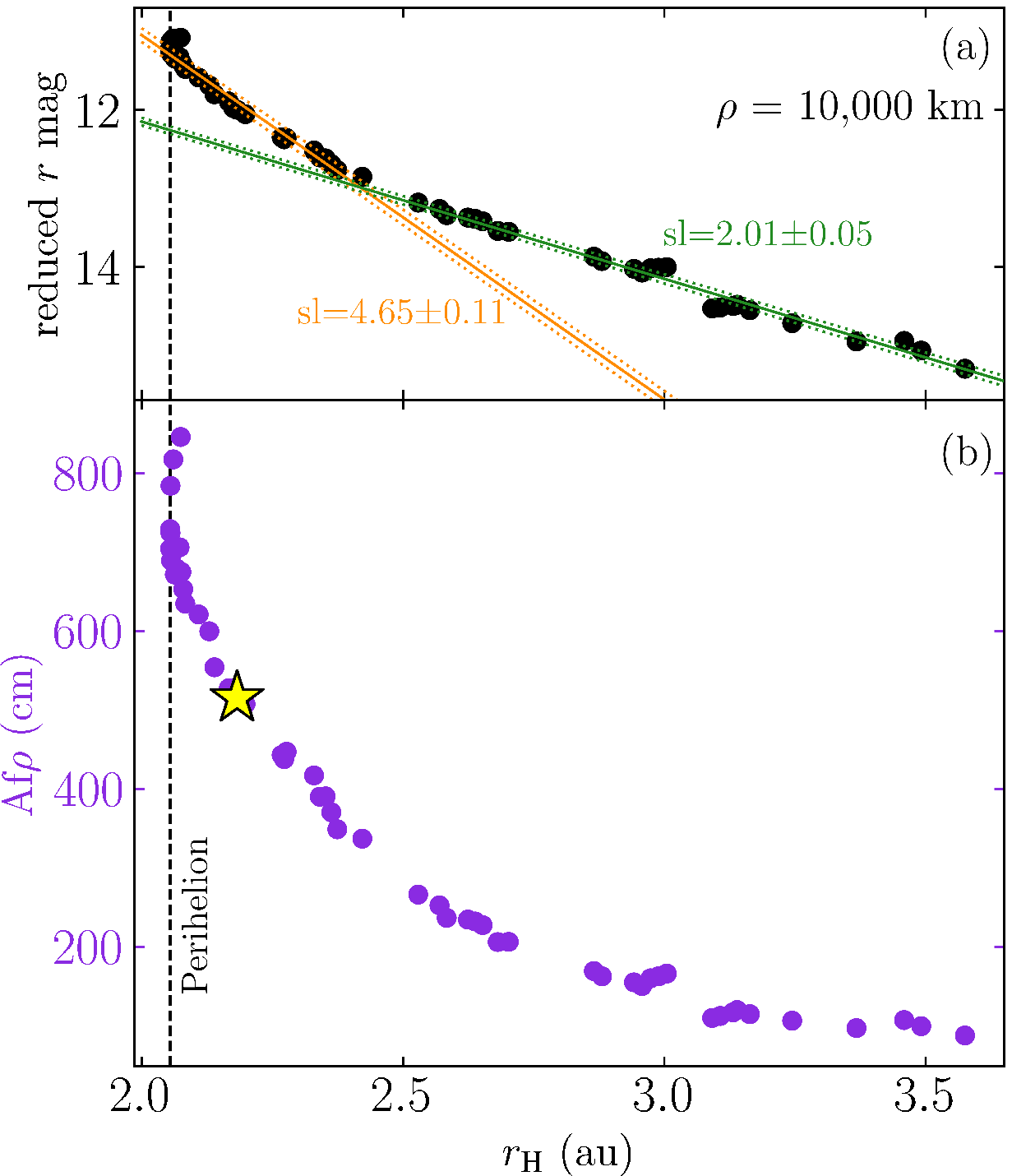}
\caption{Secular evolution of the reduced $r$-band magnitude (Panel a) and $Af\rho$ parameter (Panel b) as a function of heliocentric distance $r_{\rm H}$. All points correspond to the photometric parameters measured from the aperture size of 10 000 km from the comet nucleus. The star symbol in Panel b denotes the time of our spectropolarimetric observation (UT 2021 June 26.2). The vertical dashed line indicates the perihelion distance at $r_{\rm H}$ = 2.055 au. An increase in the brightness of T2 accelerates around 2.4 au from the Sun. A linear least-square fit to the heliocentric evolution of the reduced $r$ magnitudes gives different slopes (`sl' in panel a) before and after the acceleration.
}
\label{Figap02}
\end{figure}

Secular evolution of cometary activity as a comet approaches the Sun provides insight into how much volatile ices surface retains and thus the degree of processing of the surface layer \citep{Gundlach2015,Gundlach2020}. To put our near-infrared spectropolarimetric observation in a broader context, we utilized the $r$-band broadband imaging data of T2 from the Zwicky Transient Facility (ZTF) archive \citep{Masci2019}\footnote{\url{https://irsa.ipac.caltech.edu/applications/ztf/}}. The data cover heliocentric distances $r_{\rm H}$ ranging from 3.575 au to 2.075 au between UT 2020 October 23 and UT 2021 August 04, with the comet passing perihelion at $q$ = 2.055 au on UT 2021 July 11.1\footnote{\url{https://ssd.jpl.nasa.gov/horizons/app.html#/}}. ZTF is a 1.2-m diameter time-domain survey telescope, scanning the entire northern visible sky with a pixel scale of 1\farcs0 pixel$^{\rm -1}$ and a field of view (FoV) of 47 deg$^{\rm 2}$ to a 5-$\sigma$ median magnitude limit of $r$ $\sim$ 21.012 mags\footnote{\url{https://www.ztf.caltech.edu/ztf-camera.html}}. Since ZTF applies a unified integration time of 30 seconds \citep{Masci2019}, we made use of $r$-band data that generally displays the best signal-to-noise ratio for cometary dust. The archive provides images that are already pre-processed with bias removal, dark subtractions, and flattening. Figure \ref{Figap01} summarizes the change in the coma morphology of T2 during this period.

T2 became bright enough in the ZTF images to conduct aperture photometry on UT 2020 October 23 at $r_{\rm H}$ $\sim$ 3.6 au and began to show an extended coma signal around 3.1 au from the Sun. Throughout the apparition, T2 lacked a significant dust tail but showed a modest elongation of the dust coma (Fig. \ref{Figap01}). The oval-shaped coma containing most of the light shows no preference in the elongation direction.  In the absence of significant coma features whose signal exceeds the background uncertainty, the coma morphology in the visible light supports the nearly tailless feature of T2 in the near-infrared (Fig. \ref{Fig01}).

The evolution of the photometric parameters of T2 is shown in Figure \ref{Figap02}. $r$-band magnitudes of T2 were measured with a fixed aperture size of 10 000 km from the comet center and then corrected by differential magnitudes of background stars by comparing the instrumental magnitude of the comet with the star magnitude provided by the SDSS-12 catalog \citep{Alam2015}. The conversion between magnitudes in different catalogs was made based on the transformation parameters provided by \citet{Tonry2012} and \citet{Medford2020}. Reduced $r$-band magnitude as a function of heliocentric distance $r_{\rm H}$, $m_{\rm r}(r_{\rm H})$, was derived by correcting the effect of varying geocentric distance and phase angle over the period as
\begin{equation}
m_{\rm r}(r_{\rm H}) = m_{\rm r}(r_{\rm H}, \Delta, \alpha) - 5{\rm log}_{\rm 10}(\Delta) - 2.5{\rm log}_{\rm 10}(\Phi(\alpha))
~,
\label{eq:eq2}
\end{equation}
\noindent where m$_{\rm r}(r_{\rm H}, \Delta, \alpha)$ is the apparent magnitude, which is corrected from the differential photometry using the background star catalog and $\Phi(\alpha)$ is the phase function of the coma dust. We adopted a commonly used empirical scattering phase function (2.5 log$_{\rm 10}$($\Phi(\alpha)$) = $b\alpha$), where the phase coefficient of $b$ = 0.035 mag deg$^{\rm -1}$ was assumed (e.g. \citealt{Lamy2004}). Given the apparent magnitude, we further derived the so-called Af$\rho$ parameter (A is the albedo of dust particles and f is their packing density within the aperture radius of $\rho$; \citealt{A'Hearn1984}), a proxy of the dust production rate \citep{Fulle2022}, using Eq. 8 in \citet{Kwon2019}, where $\rho$ = 10 000 km throughout the analysis.

At around 2.4 au, the slope of brightening in the reduced $r$-band magnitude begins to steepen more than two times higher than at further distances (Fig. \ref{Figap02}). The $Af\rho$ parameter accordingly exhibits an acceleration around that point, indicative of the discontinuous dust ejection into the coma. The dust production rate, approximated by the $Af\rho$ parameter, of T2 is $\gtrsim$100 times weaker than other OCCs at similar $r_{\rm H}$ \citep{Mazzotta Epifani2016,Garcia2020,Fulle2022}.

Different types of ice particles sublimate at different temperature environments and thus dominate cometary activity at distinctive distance regimes from the Sun. Supervolatile ice (particularly CO$_{\rm 2}$ ice) sublimation is significant until $\sim$4 au, followed by the regime where water ice starts to dominate the dust ejection at $\sim$2.5--2.7 au
\citep{Blum2014,Gundlach2015,Bauer2015,Gundlach2020}.  The observed transitional point of T2 at $\sim$2.4 au would signify the onset point of water ice sublimation-driven dust activity of the comet. The very low-activity level outside the transition point implies a dearth of supervolatile ices near the surface dust layer and/or that the upper dust layer through which gas molecules diffuse outward is more consolidated compared to those of other active comets. Both cases indicate that there is a processed surface environment on T2's nucleus. These surface conditions seem to be far different from those of comets that have just completed their gravitational aggregation in the protoplanetary system (\citealt{Blum2022} and references therein) or those of comets in early evolutionary phases where their surfaces deplete inherent ice by sublimation but have not yet undergone significant compression that can preserve highly fluffy structures
\citep{Skorov2012,Poch2016a,Poch2016b}.

Putting all the results together, T2 appears to share several aspects with so-called Manx comets: not only similar locations in orbital space (Fig. \ref{Fig06}) but also their relatively low activity in scattered light than active dust-rich OCCs, (nearly) tailless coma morphology \citep{Meech2016,Piro2021}, and the existence of a discontinuous, transitional point of the dust ejection in their inbound orbits (e.g. \citealt{Molnar-Bufanda2019}). All the observations suggest the paucity of highly porous (or ensuing $\sim$0.1--1 $\mu$m-sized small) dust particles in their pre-perihelion comae. Observations of low-activity OCCs in thermal emission (particularly in the mid-infrared of 5--20 $\mu$m) will provide independent information on how the internal structure and composition of their dust constituents \citep{Gehrz1992} differentiate from active OCCs. This will help reveal diversities embedded in the formation and subsequent evolutionary history of comets consisting of the present-day Oort cloud.

\section{Ancillary information for the comets used in Figure \ref{Fig06}\label{sec:app2}}

\begin{table*}[!b]
\small
\centering
\caption{Properties of the comets used in Figure \ref{Fig06} and their references}
\vskip-1ex
\begin{tabular}{l|l|cccc|c|cc}
\toprule
\hline
\multirow{2}{*}{Dyn. Type} & \multirow{2}{*}{Comet Name} & 1/$a_{\rm ori}$ & 1/$a_{\rm osc}$ & $q_{\rm osc}$ & $i_{\rm osc}$ & Obs. & \multicolumn{2}{c}{References}\\
\cline{8-9}
 & & (10$^{\rm -6}$ au$^{\rm -1}$) & (10$^{\rm -6}$ au$^{\rm -1}$) & (au) & (\degree) & Type & \multicolumn{1}{c|}{Dyn. Type} & Obs. Type \\ 
\hline
\multirow{17}{*}{Oort-spike comets} & C/1940 R2 (Cunningham) & $-$60.010 & $-$1380.070 & 0.368 & 49.895 & I & \multicolumn{1}{c|}{(1)} & (4) \\
 & C/1956 R1 (Arend-Roland) & 19.740 & $-$784.895 & 0.316 & 119.944 & II & \multicolumn{1}{c|}{(2)} & (5) \\
 & C/1973 E1 (Kohoutek) & 19.930 & $-$54.835 & 0.142 & 14.304 & II & \multicolumn{1}{c|}{(2)} & (6), (7) \\
 & C/1989 Q1 (OLR) & 42.900 & $-$86.684 & 0.642 & 90.146 & I & \multicolumn{1}{c|}{(1)} & (7), (8) \\
 & C/1989 X1 (Austin) & 40.840 & $-$652.084 & 0.350 & 58.956 & I & \multicolumn{1}{c|}{(1)} & (7), (9) \\
 & C/1993 A1 (Mueller) & 61.840 & $-$950.152 & 1.938 & 124.878 & II & \multicolumn{1}{c|}{(1)} & (7), (10) \\
 & C/1999 S4 (LINEAR) & $-$54.510 & $-$141.544 & 0.765 & 149.385 & II & \multicolumn{1}{c|}{(2)} & (11) \\ 
 & C/2001 Q4 (NEAT) & 60.560 & $-$715.260 & 0.962 & 99.643 & II & \multicolumn{1}{c|}{(1)} & (9), (12) \\
 & C/2002 T7 (LINEAR) & 25.650 & $-$791.285 & 0.615 & 160.583 & II & \multicolumn{1}{c|}{(1)} &  (13) \\
 & C/2003 K4 (LINEAR) & 31.390 & $-$293.766 & 1.024 & 134.253 & I & \multicolumn{1}{c|}{(1)} & (14), (15) \\
 & C/2007 N3 (Lulin) & 29.310 & 13.510 & 1.212 & 178.374 & I & \multicolumn{1}{c|}{(1)} & (16), (17) \\
 & C/2007 W1 (Boattini) & $-$29.120 & $-$0.0003 & 0.850 & 9.890 & I & \multicolumn{1}{c|}{(1)} & (15), (18) \\
 & C/2011 L4 (PANSTARRS) & 29.480 & $-$108.664 & 0.302 & 84.208 & II & \multicolumn{1}{c|}{(2)} & (17) \\
 & C/2012 S1 (ISON) & $-$36.060 & $-$408.744 & 0.013 & 62.163 & I & \multicolumn{1}{c|}{(1)} & (19), (20) \\
 & C/2013 US10 (Catalina) & 52.960 & $-$340.367 & 0.823 & 148.878 & I & \multicolumn{1}{c|}{(2)} & (21), (22) \\
 & C/2013 V1 (Boattini) & 27.670 & $-$859.225 & 1.661 & 65.310 & II & \multicolumn{1}{c|}{(2)} & (23) \\
 & C/2017 K2 (PANSTARRS)$^{\star}$ & 35.490 & $-$226.514 & 1.800 & 87.543 & II & \multicolumn{1}{c|}{(1)} & (24) \\
\hline
\multirow{26}{*}{Long-period comets} & C/1957 P1 (Mrkos) & 2000.790 & 1865.108 & 0.355 & 93.957 & II & \multicolumn{1}{c|}{(2)} & (25), (26) \\
 & C/1974 C1 (Bradfield) & 581.990 & 528.562 & 0.503 & 61.285 & II & \multicolumn{1}{c|}{(2)} & (27), (28) \\
 & C/1975 N1 (KBM) & 816.400 & $-$215.618 & 0.426 & 80.781 & I & \multicolumn{1}{c|}{(2)} & (29), (30) \\
 & C/1983 H1 (IAA) & 10680.070 & 10200.506 & 0.991 & 73.251 & I & \multicolumn{1}{c|}{(2)} & (7) \\
 & C/1987 P1 (Bradfield) & 6379.200 & 6056.545 & 0.869 & 34.088 & II & \multicolumn{1}{c|}{(2)} & (30), (31) \\
 & C/1988 A1 (Liller) & 4880.400 & 4084.722 & 0.8841 & 73.322 & II & \multicolumn{1}{c|}{(2)} & (28), (32)\\
 & C/1990 K1 (Levy) & 113.630 & $-$620.174 & 0.939 & 131.584 & II & \multicolumn{1}{c|}{(1)} & (30), (33) \\
 & C/1991 T2 (Shoemaker-Levy) & 936.350 & 166.289 & 0.836 & 113.497 & II & \multicolumn{1}{c|}{(2)} & (28) \\
 & C/1992 F1 (Tanaka-Machholz) & 3422.100 & 3189.030 & 1.262 & 79.292 & II & \multicolumn{1}{c|}{(2)} & (34) \\
 & C/1995 O1 (Hale-Bopp) & 3838.980 & 5492.937 & 0.917 & 89.217 & II & \multicolumn{1}{c|}{(2)} & (35), (36), (37) \\
 & C/1996 B2 (Hyakutake) & 1546.310 & 470.642 & 0.230 & 124.922 & II & \multicolumn{1}{c|}{(2)} & (38), (39) \\
 & C/1996 Q1 (Tabur) & 1876.000 & 1652.840 & 0.840 & 73.356 & I & \multicolumn{1}{c|}{(3)} & (40) \\
 & C/2000 WM1 (LINEAR) & 532.510 & $-$437.171 & 0.555 & 72.550 & II & \multicolumn{1}{c|}{(2)} & (41), (42) \\
 & C/2001 A2 (LINEAR) & 1110.600 & 395.174 & 0.779 & 36.487 & I & \multicolumn{1}{c|}{(2)} & (43), (44) \\
 & C/2004 Q2 (Machholz) & 419.150 & 413.831 & 1.205 & 38.589 & I & \multicolumn{1}{c|}{(2)} & (45) \\
 & C/2009 P1 (Garradd) & 421.010 & $-$643.289 & 1.551 & 106.177 & II & \multicolumn{1}{c|}{(2)} & (46) \\ 
 & C/2012 L2 (LINEAR) & 2540.640 & 1767.598 & 1.509 & 70.981 & II & \multicolumn{1}{c|}{(2)} & (47) \\
 & C/2013 R1 (Lovejoy) & 2722.000 & 1939.842 & 0.812 & 64.041 & II & \multicolumn{1}{c|}{(3)} & (48) \\
 & C/2013 UQ4 (Catalina) & 17209.130 & 16509.594 & 1.081 & 145.259 & II & \multicolumn{1}{c|}{(2)} & (49) \\
 & C/2020 T2 (Palomar) & 2778.000 & 3193.207 & 2.055 & 27.873 & I & \multicolumn{1}{c|}{(2)} & This Study \\
\cline{2-9}
 & C/2014 S3 (PANSTARRS)$^{\dagger}$ & 11809.130 & 10879.747 & 2.050 & 169.321 & I & \multicolumn{1}{c|}{(2)} & (50) \\
 & C/2013 P2 (PANSTARRS)$^{\dagger}$ & 1284.720 & 387.375 & 2.835 & 125.532 & I & \multicolumn{1}{c|}{(2)} & (51) \\
 & A/2018 V3$^{\dagger}$ & 8557.000 & 8234.440 & 1.340 & 164.977 & I & \multicolumn{1}{c|}{(2)} & (52) \\
 & C/2002 CE10 (LINEAR)$^{\dagger}$ & 101736.630 & 101875.559 & 2.047 & 145.459 & I & \multicolumn{1}{c|}{(2)} & (53) \\
 & C/2016 VZ18 (PANSTARRS)$^{\dagger}$ & 5090.920 & 5138.714 & 0.910 & 24.036 & I & \multicolumn{1}{c|}{(2)} & (54) \\
 & C/2017 O1 (ASASSN)$^{\dagger}$ & 2477.000 & 2385.485 & 1.499 & 39.848 & I & \multicolumn{1}{c|}{(2)} & (55) \\
\hline
\bottomrule
\end{tabular}
\tablefoot{Numbered references indicate (1) the Polish Catalogue of Near-Parabolic Comets (\url{http://ssdp.cbk.waw.pl/LPCs/near_parabolic_comets_catalogue.html}); (2) Minor Planet Centre Database Search (\url{https://www.minorplanetcenter.net/db_search}); (3) the Nakano Note (\url{https://www.oaa.gr.jp/~oaacs/nk.htm}); (4) \citet{Ohman1941}; (5) \citet{Finson1968b}; (6) \citet{A'Hearn1975}; (7) \citet{Hanner1994}; (8) \citet{Rosenbush1994}; (9) \citet{Kwon2021}; (10) \citet{Levasseur-Regourd1996}; (11) \citet{Kiselev2002}; (12) \citet{Wooden2004}; (13) \citet{Rosenbush2006}; (14) \citet{Woodward2004}; (15) \citet{Velichko2012}; (16) \citet{Woodward2011}; (17) \citet{Choudhury2015}; (18) \citet{Gibb2012}; (19) \citet{Wooden2014}; (20) \citet{Snios2016}; (21) \citet{Kwon2017}; (22) \citet{Woodward2021}; (23) \citet{Roy2015a}; (24) \citet{Zhang2022}; (25) \citet{Kearns1958}; (26) \citet{Martel1960}; (27) \citet{Ney1974}; (28) \citet{Kiselev2015}; (29) \citet{Ney1982}; (30) \citet{Chernova1993}; (31) \citet{Hanner1990}; (32) \citet{Turner1999}; (33) \citet{Lynch1992}; (34) \citet{Geyer1996}; (35) \citet{Kiselev1997}; (36) \citet{Mason2001}; (37) \citet{Harker2002}; (38) \citet{Kiselev1998}; (39) \citet{Mason1998}; (40) \citet{Harker1999}; (41) \citet{Joshi2003}; (42) \citet{Kelley2004}; (43) \citet{Rosenbush2002}; (44) \citet{Furusho2003}; (45) \citet{Velichko2005}; (46) \citet{Hadamcik2014}; (47) \citet{Roy2015b}; (48) \citet{Borisov2015}; (49) \citet{Ivanova2017}; (50) \citet{Meech2016}; (51) \citet{Meech2014}; (52) \citet{Piro2021}; (53) \citet{Sekiguchi2018}; (54) \citet{Bufanda2020}; (55) \citet{Brinkman2020}.
$^\dagger$ Manxes or Manx candidates. $^{\star}$Although the observations conducted within 3.1 au from the Sun have not been published, several studies observing the comet outside the water ice line (e.g. \citealt{Yang2021,Fulle2022}) and optical images from amateurs inside the line (e.g. \url{https://cometografia.es/2017k2-panstarrs-2022-07-15/#more-14767}) support its membership in Type II. We will present more detailed narrowband imaging and polarimetric aspects of its dust in our future work.}
\label{ta1}
\vskip-1ex
\end{table*}

\end{appendix}
\end{document}